# Direct Orientation Contrast Imaging of Anti-Phase Domains on III-V Materials Using Scanning Electron Microscopy

Brieg Le Corre[1,2], Clothilde Grenèche[1], Rozenn Bernard[1], Tony Rohel[1], Antoine Létoublon[1], Wijden Khelifi[1], Julie Le Pouliquen[1], Arnaud Grisard[3], Sylvain Combrié[3], Bruno Gérard[4], Abdelmounaim Harouri[2], Luc Le Gratiet[2], Grégoire Beaudoin[2], Konstantinos Pantzas[2], Isabelle Sagnes[2], Oliver Skibitzki[5], Gilles Patriarche[2], Charles Cornet[1] and Yoan Léger[1,*]

[1]Univ Rennes, INSA Rennes, CNRS, Institut FOTON - UMR 6082, F-35000 Rennes, France
[2]Centre de Nanosciences et de Nanotechnologies, CNRS, 91120 Palaiseau, France
[3]Thales Research and Technology, 91767 Palaiseau, France
[4]III-V Lab, 91767 Palaiseau, France
[5]IHP-Leibniz Institut für Innovative Mikroelektronik, 15236 Frankfurt (Oder), Germany

**ABSTRACT**. Direct orientation contrast imaging of zinc-blende III-V materials is studied using scanning electron microscopy. A quantitative approach is taken using a 3 µm thick orientation-patterned GaP grown on GaAs sample, studying the anti-phase domain contrast with respect to the electron beam energy and the tilt angle. A qualitative approach is taken for III-V grown on non-polar materials with and without chemical mechanical polishing. Finally, a processing of the acquired image for GaP on Si reveals in plane preferential anti-phase boundaries.

## I. INTRODUCTION

Crystallographic textures in III-V materials are attracting more and more whether it arises naturally, for instance in III-Vs heteroepitaxy on silicon (Si) or germanium (Ge) (GaSb/Si [1], GaAs/Si [2,3], GaAs/Ge [4–6], GaP/Si [7–11]) or by design, in orientation-patterned (OP) III-Vs for nonlinear optics (OP-GaSb [12], OP-GaAs [13], OP-GaP [14–16], OP-AlGaAs [5] and OP-GaAsP [17]).

In the first case, the growth of polar materials on non-polar substrates, such as III-Vs on Si or on Ge, naturally leads to randomly distributed texturing in the form of inverted orientation. These anti-phase domains (APDs) and their boundaries (APBs) are often considered as defects, particularly for active optoelectronic devices, such as emitters on silicon [18] or energy devices [19], where they constitute short-circuit paths. Several epitaxial strategies have been used to mitigate their detrimental properties in e.g. lasers [20,21], or in photovoltaic solar cells [22]. Additionally, recent proposals were made to use their electronic properties for solar hydrogen production [23]. In the second case, in OP-III-Vs, the crystal polarity is flipped by design along the <110> axis, forming rectangular domains of inverted orientation to ensure quasi-phase matching of second order nonlinear optical processes in waves propagating along <110>. In both cases, the characterization and identification of boundaries/domains, their atomic configuration and their behavior during the growth has been the subject of intense research in recent years [8,24–26].

So far, the characterization of these inverted orientations has been mainly achieved by transmission electron microscopy (TEM) imaging [4,27], or scanning electron microscopy (SEM) imaging associated with APB-selective etching [28–30], both having the drawback of being destructive techniques. Cross- and plane-view Scanning Tunneling Microscopy (STM) have been also proposed with the constraint of preserving the surface of oxidation using engineered capping layers or in-situ cleaving [31,32].

Therefore, the development of a fast, non-destructive, qualitative and quantitative imaging and characterization technique for III-V semiconductors crystallographic inverted orientations is a crucial resource in the development of advanced photonic and energy devices based on these materials.

To achieve this objective, orientation characterization via SEM stands out as a particularly effective method. Traditionally, this approach relied on electron backscatter diffraction (EBSD), a technique that requires, an expensive bespoke detector, advanced understanding of electron diffraction physics, and demanding experimental procedures [33,34]. Over the past decade, however, the widespread adoption of in-lens detectors has enabled the observation of crystallographic textures without the need for EBSD technology, as demonstrated on electropolished metals in [35] where, to our knowledge, the term of "orientation contrast imaging" is used for the first time when talking about in-lens detectors imaging. It was also subsequently demonstrated on SiC in [36] and on copper in [37].

In this work, a relevant preliminary method for in-depth analyses such as EBSD and TEM is studied. We will refer to our method of observing orientation contrast in SEM as direct orientation contrast imaging (DOCI), and present its use applied to III-V/Si epilayers and OP-III-V materials. We first review the theory behind DOCI to identify the experimental conditions for the observation of inverted domains on III-V/Si and OP-III-V materials. We then we apply this method to OP-GaP to quantitatively assess the contrast of DOCI depending on experimental conditions. The third section of the article evaluates whether the DOCI technique is applicable on various flat or rough III-V semiconductors grown on Si and the last part of the article illustrates how the DOCI technique can be used to give the first quantitative characterizations of GaP/Si epilayers, revealing preferential in-plane orientations of APBs.

## II. METHODS
### A. Principle of DOCI

*Contact author: yoan.leger@insa-rennes.fr

In SEM imaging there are essentially two ways to access crystallographic orientation information, EBSD and electron channeling patterns (ECP)/electron channeling contrast imaging (ECCI). Early experimental observations by Coates (1967) [38] and Schulson & van Essen (1969) [39] showed that bulk crystals produce channeling patterns if the incident beam orientation varies in SEM. The theoretical framework relies on dynamical diffraction theory, extended to SEM by Spencer, Humphreys & Hirsch (1972) [40] by providing a forward/backward scattering model in which the incident electrons are described as Bloch waves whose propagation and scattering depend on crystal orientation. If aligned with specific lattice directions, these waves channel deeply, reducing scattering; whereas misalignment enhances backscattering, producing the contrast observed in ECP/ECCI.

EBSD, on the other hand, originates from a different electron population, namely high-energy inelastically scattered electrons, which subsequently undergo exit-wave Bragg diffraction to form Kikuchi patterns [41,42]. The geometric structure of the bands in EBSD and ECP is fundamentally similar, as both reflect the same underlying Bragg conditions. However, the contrast is reversed: in EBSD, band maxima correspond to orientations that enhance diffracted exit intensity, whereas in ECP/ECCI the backscattering signal is reduced when the primary electrons meet channeling directions. Thus, ECP can be viewed as an inverse-contrast analogue of EBSD Kikuchi patterns. Furthermore, EBSD provides images defined along the detection directions, whereas ECP/ECCI images are related to the excitation directions. Therefore, ECP Kikuchi-like patterns can be obtained in SEM using rocking beam [43] or rocking stage [44,45] configurations, and yield complementary but structurally related Kikuchi-band information to EBSD.

In the present case, DOCI derives precisely from the difference in channeling between two opposite orientations of a non-centrosymmetric material.

*Contact author: yoan.leger@insa-rennes.fr

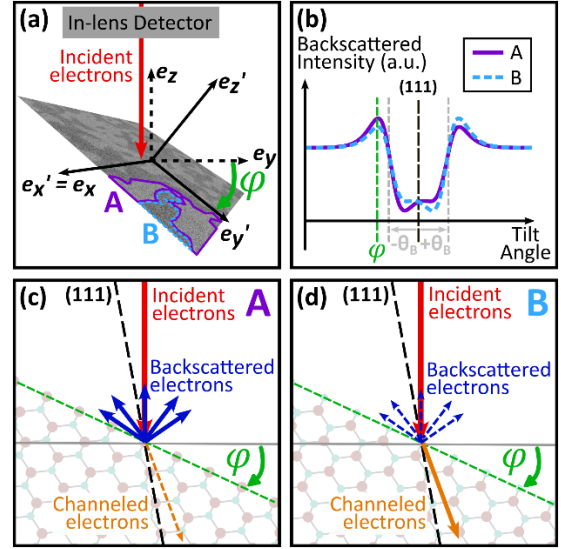

FIG 1. (a) Representation in the SEM chamber when applying a tilt angle $\varphi$ on a sample presenting domains of opposite orientation (A and B). (b) Schematic representation of the backscattered intensity on domain A (in purple line) and domain B (in blue dashed line) by considering only first-order diffraction around (111) near Bragg diffraction conditions. (c) and (d) illustrations of the backscattered intensity of domain A and domain B respectively, following (b) at tilt angle $\varphi$.

Figure 1 depicts this schematically. Figure 1 (a) represents the sample in the SEM chamber, where $(e_x, e_y, e_z)$ is the initial referential when no rotation and no tilt are applied. $(e'_x, e'_y, e'_z) = ([1\bar{1}0], [110], [001])$ denotes the sample referential after a tilt angle $\varphi$ is applied without azimuthal rotation, i.e. the sample tilting is done around $[1\bar{1}0]$ axis. Figure 1 (b) illustrates the backscattered intensity with respect to the tilt angle $\varphi$, around an anion or cation diffractive plane (here (111)), depending on the domain the beam is focused in (A or B, which have opposite orientations). Figure 1 (c) and (d) portray the backscattering and channeling of incident electrons at tilt angle $\varphi$, for domain A and B respectively. As illustrated in Figures 1 (b), (c) and (d), domain A will induce more backscattering and less channeling than domain B at tilt angle $\varphi$ explaining the resulting orientation contrast between A and B.

### B. Sample preparations and experimental procedures

In total, seven samples were investigated, including III-Vs grown on Si (both as-grown and with a polished surface), and OP-GaP samples. These samples represent both natural or artificially-designed antiphase domains, in a variety of materials (GaP, GaAs, GaPSb and InGaP). First, a 3 µm-thick OP-GaP sample was prepared by following the process described in [16]. This sample then went through a chemical-mechanical polishing (CMP) step to reduce the surface roughness down to 1.6 nm RMS. Finally, 10 µm x 5 µm openings were created in a plasma-deposited SiN mask by a lithography-etching step in order to conduct the quantitative study of DOCI on well-defined positions. Second, three III-V/Si samples were grown by Molecular Beam Epitaxy (MBE) for assessing the chemical sensitivity of the DOCI technique. These three samples were polished using CMP, reducing roughness to less than 1 nm RMS. They are respectively composed of GaP (400 nm-thick), $GaP_{0.4}Sb_{0.6}$ (1 µm-thick) and GaAs (1 µm-thick) grown on Si substrates following the experimental details given in [46], [47] and [48]. Third, two III-V/Si samples were grown by MBE for evaluating the robustness of DOCI against roughness. These two samples remained as-grown and were not polished. A 1 µm-thick GaP/Si sample was elaborated following the experimental conditions given in [46]. A 270nm-thick $In_{0.3}Ga_{0.7}P$ layer was also grown on a $Si_{0.43}Ge_{0.57}$ 500 nm-thick metamorphic buffer deposited on a Si substrate, following the approach described in [49]. Finally, a GaP epilayer (500nm-thick) was grown on Si by MBE and polished by CMP, to illustrate how quantitative, statistical information on domains and boundaries can be obtained from DOCI data.

Moreover, DOCI measurements were carried out in two different microscopes equipped with different in-lens detector, to demonstrate that the technique can be deployed in any SEM as long as it possesses an in-lens detector. The first is a Verios G4 HP Thermo Fisher Scientific SEM using either the "Through Lens Detector" (TLD SE (secondary electrons) or BSE (backscattered electrons) modes), later referred to as TLD measurements. The second is an Apreo 2C Thermo Fisher Scientific SEM using the T1 in-lens detector, later referred to as T1 measurements. The samples were placed so that their [110] axis is parallel to the tilt plane of the microscope. Then images were taken for tilt angle varying between 0 and 45° using different beam energy. The working distance was fixed at 5 mm for TLD measurements and 10 mm for T1 measurements.

First, to quantitatively assess the performance of DOCI, images were systematically recorded across the whole tilt range onto a single opening of the SiN mask of the OP-GaP sample, for each e-beam probe condition that were used: (5 keV, 100 pA), (10 keV, 200 pA) and (20 keV, 200 pA). Long measurements on the same area can lead to surface carbon contamination and one must be vigilant about the decrease in contrast caused by this phenomenon.

A post-processing step was then employed to compensate for the shift in the overall intensity of the image due to the tilt dependence of the detector sensitivity, related to its position and finite numerical aperture, and progressive surface carbon contamination. For this purpose, an image with a contrast sufficiently high to distinguish the domains. In this image, one-region of interest (ROI) was defined in each domain. These ROIs were placed in relation to the edge of the opening, which has the advantage of being easy to detect whatever the observation conditions used. Figure 2 shows an example of this procedure, with four ROIs of equal size placed at different positions. Note that the size of these ROIs decreases with increasing tilt angle due to image perspective.

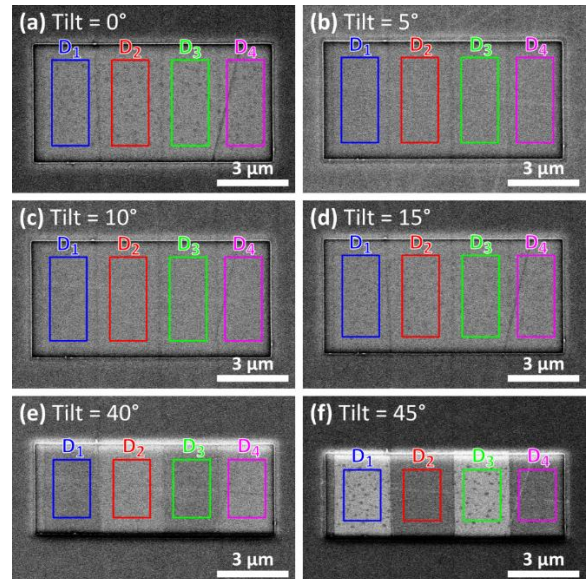

FIG 2. Representation of the four selected ROIs: D1 in blue, D2 in red, D3 in green and D4 in magenta. After post-processing of T1 images taken at 10 keV, for tilt angles of: (a) 0°, (b) 5°, (c) 10°, (d) 15°, (e) 40° and (f) 45°.

As the tilt angle varies, global contrast and brightness may change between two acquisitions. The image intensity is therefore renormalized through a linear transformation (LT) for each angle to make the orientation contrast from one angle to the other comparable. A similar intensity renormalization was also carried out in the case of III-V samples grown on Si which have undergone a CMP step. Additional

*Contact author: yoan.leger@insa-rennes.fr

information on this procedure is provided in the Supplementary Materials.

The orientation contrast is estimated as follows: Two values of contrasts are calculated: $C_{12}$ (between D1 and D2) and $C_{34}$ (between D3 and D4) as $C_{kj} = \frac{\overline{I_{LT,k}} - \overline{I_{LT,j}}}{\overline{I_{LT,k}} + \overline{I_{LT,j}}} \times 100$. The overall orientation contrast is then calculated as the average value of $C_{12}$ and $C_{34}$. Uncertainties in the overall contrast is given as a 70%-confidence interval around the calculated contrast value, as shown in Figure 3. Additional information about the uncertainty and how it is calculated is available in the Supplementary Materials.

## III. RESULTS AND DISCUSSION
### A. OP-GaP observations

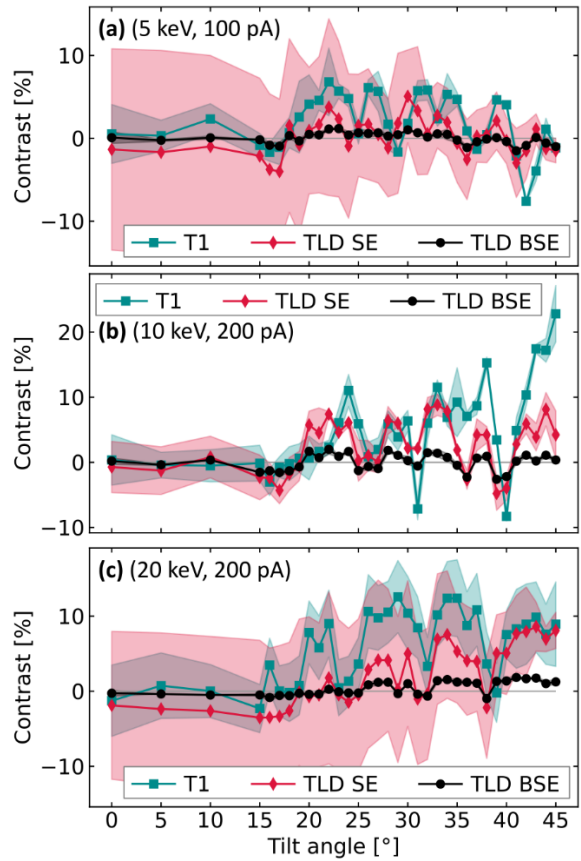

FIG 3. Orientation contrast evolution, with their 70%-confidence interval, with respect to the tilt angle for beam energy of: (a) 5 keV, (b) 10 keV and (c) 20 keV; using T1, TLD-SE mode and TLD-BSE mode represented in dark cyan, crimson and black respectively.

The orientation contrast varies with the tilt angle and incident beam energy as presented in Figure 3. Qualitatively, the contrasts recorded on the two microscopes follow the same trends with decrease and even reversal of the contrast at specific angles (e.g. for 5 keV: 15-17°, 27-28°, 36-37° and 41-42°). These variations are very strong in some cases: up to 30% peak-to-peak on T1 measurements at 10 keV. Contrast reversal can be appreciated on the SEM images presented in Figure 2(e) and (f). A slight mismatch of ~1° in tilt angle at which some contrast variations occur is observed between the TLD measurements and the T1 measurements. This is attributed to the difference in calibration of the sample stages of the two microscopes and positioning reproducibility. The angular dependence of the contrast is complex. In spite of this, conditions can be found to optimize the contrast for any beam energy configuration.

Apart from the TLD-SE measurements at 5 and 20 keV (discussed in the following), the recorded orientation contrast is close to 0 at small tilt angles (below 5°) for all detectors and all e-beam configurations and strongly increases above 15°. A very high contrast is obtained for a beam energy of 20 keV and a tilt angle of 34° (close to (111) plane at Bragg angle, 35.3° – 0.78° = 34.52°) on all detectors. For T1 measurements, an even higher contrast is obtained when the beam energy is 10 keV and the sample tilt angle 45° (close to (559) plane at Bragg angle, 51.8° – 7.36° = 44.44°).

To reflect the complexity of the angular dependence of the orientation contrast that we should expect, table I summarizes the diffraction conditions of the main polar planes involved during tilting the sample from 0° (incidence parallel to [001] axis) towards the [110] axis. It is also worth mentioning that the orientation contrast does not necessarily show up similarly in the diffraction patterns for all these planes [50]. We should also expect contributions from nonpolar planes, contributing to an offset on the signal and contributions from other planes that are not perpendicular to the [1-10] axis.

Based on Table I, one may think that the contrast observed in the TLD measurements (SE mode) at 5keV and 20keV at very low angles, can be attributed to diffraction onto the (551) and (771) planes respectively. However, we do not observe the same effect on the BSE mode measurement (recorded on the same mask opening). Experimental artefacts such as residual topology on a domain (e.g. CMP scratch, see Figure S1 in Supplementary Materials) can also result in an artificial offset on the measured contrast. The SE mode of the TLD being more sensitive to the surface than the BSE mode, it is prone to suffer strongly from this kind of artifact. The absolute value of the contrast or the precise position of contrast reversal can thus hardly be analyzed quantitatively. It is more relevant to focus on relative changes in the orientation contrast from one angle to the other to identify satisfying observation conditions.

*Contact author: yoan.leger@insa-rennes.fr

Finally, one can notice that despite a similar behavior of the TLD measurements in SE and BSE modes, the latter show much less contrast than the former. This is similar to what has been observed on metallic samples [35] and suggests that the T1 detector of the Apreo 2C microscope is sensitive to secondary electrons induced by backscattered electrons, rather than to backscattered electrons themselves.

TABLE I. Bragg angle, at first order, for the "first" diffractive planes of zinc-blende structure.

| Plane (hkl) | Angle from normal [hkl] to [110] axis (°) | 1st order Bragg Angle (°) | | | |
|---|---|---|---|---|---|
| | | 2 keV | 5 keV | 10 keV | 20 keV |
| (111) | 35.3 | 2.50 | 1.58 | 1.11 | 0.78 |
| (331) | 13.3 | 6.29 | 3.97 | 2.80 | 1.97 |
| (551) | 8.05 | 10.3 | 6.51 | 4.59 | 3.23 |
| (553) | 23.0 | 11.1 | 7.00 | 4.93 | 3.47 |
| (557) | 44.7 | 14.5 | 9.09 | 6.40 | 4.50 |
| (559) | 51.8 | 16.7 | 10.5 | 7.36 | 5.17 |
| (771) | 5.77 | 14.5 | 9.09 | 6.40 | 4.50 |
| (773) | 16.9 | 15.1 | 9.45 | 6.65 | 4.67 |
| (775) | 26.8 | 16.2 | 10.1 | 7.15 | 5.01 |
| (779) | 42.3 | 19.7 | 12.3 | 8.62 | 6.05 |

In conclusion, a wide range of experimental conditions for DOCI on III-V samples are available, in terms of both tilt angle and beam acceleration voltages, making the technique well adapted to both very thin films and thick samples and to a large range of electronic microscopes.

## B. III-V on Si observations
### 1. Polished samples

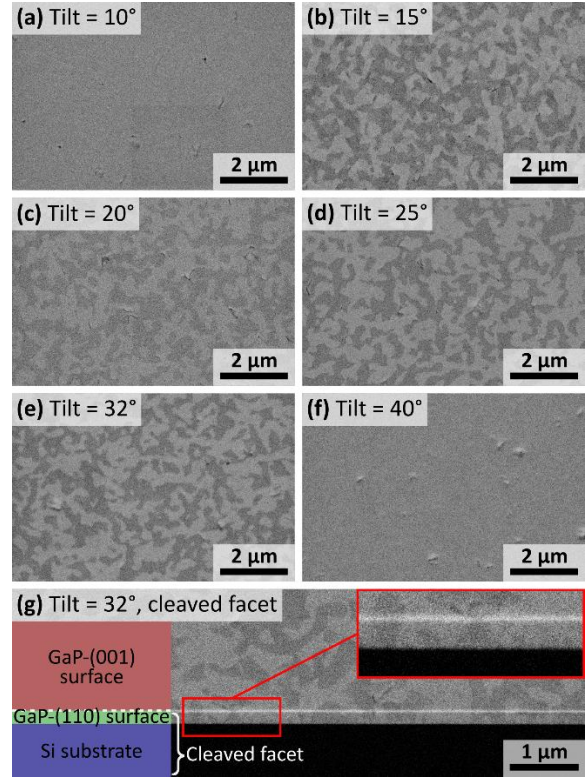

FIG 4. Orientation contrast evolution on GaP grown on Si and polished by CMP, using T1 detector at 2 keV with a working distance of 10 mm, applying tilt angle of (a) 10°, (b) 15°, (c) 20°, (d) 25°, (e) 32° and (f) 40°. (g) repeats the observation conditions used in (e) except that the image is taken at a cleaved facet of the sample; the bright line that appears is the intersection of the (001) surface with the (110) cleavage plane. The inset here shows the (110) surface of GaP.

As III-V samples grown on Si present smaller thicknesses than the one of OP-GaP, the beam energy used to analyze these samples was reduced to 2 keV with a current of 200 pA and the samples polished to prevent any influence of the surface topology. First, Figure 4 shows a selection of images recorded at specific tilt angles highlighting the orientation contrast behavior on a GaP/Si sample. A maximum contrast value is once again observed close to the Bragg condition on the (111) plane at 32° (see Table 1). It should also be noted that this contrast is also visible on the (110) cleavage plane surface, as shown in Figure 4(g) DOCI therefore provides surface-level information on the APDs on the sample as well as in-depth information if the acquisition is done on a cleaved facet. Smaller contrast values can still be observed between 15° and 32°.

These results can be straightforwardly compared to the ones of part III.A. since they are related to the same binary compound. GaP is a highly polar semiconductor with an atomic number difference $\Delta Z = 16$. It represents therefore a model system for testing DOCI, and one may wonder whether this technique is also suitable for semiconductors that have a weaker polar character. To assess this point, we tested DOCI on both $GaP_{0.4}Sb_{0.6}/Si$ ($\Delta Z = 5.6$) and GaAs/Si ($\Delta Z = 2$). Figure 5(a) shows the best orientation contrast obtained on $GaP_{0.4}Sb_{0.6}/Si$ using a tilt angle of 16°, where domains are clearly evidenced and Fig. 5(b) shows that a slight contrast can even be recorded on GaAs/Si using a tilt of 33° (see a detailed analysis in Supplementary Material).

*Contact author: yoan.leger@insa-rennes.fr

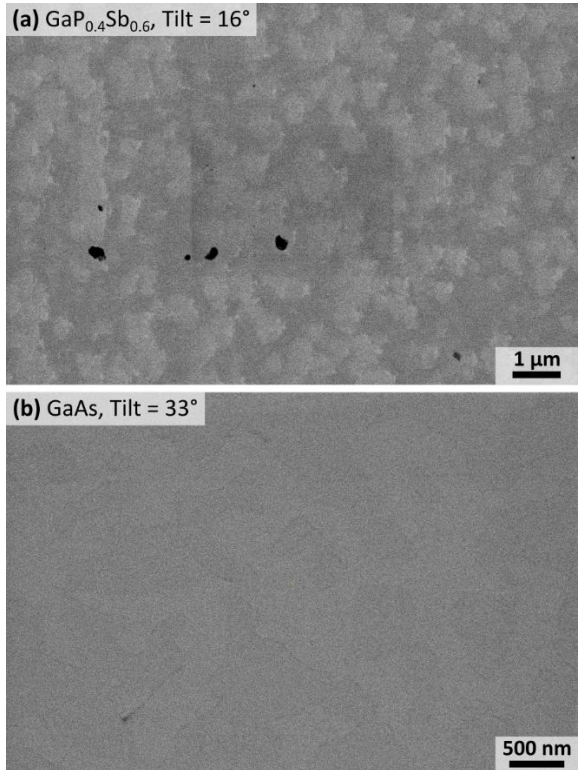

FIG 5. (a) Orientation contrast for $GaP_{0.4}Sb_{0.6}$ after CMP, 2 keV, 200 pA, tilt angle 16°. (b) Orientation contrast for GaAs/Si after CMP, 2 keV, 100 pA, tilted at 33°.

### 2. Unpolished sample observations

We conducted several observations on unpolished samples as presented in Figure 6, on GaP/Si and on $In_{0.3}Ga_{0.7}P/SiGe$ ($\Delta Z = 21.4$) using the Apreo 2C microscope. As observed in Figure 6(a) and (b), the use of a standard Everhart-Thornley detector (ETD) positioned in the microscope chamber, strongly sensitive to topology, does not allow observation of orientation contrast due to the large as-grown roughness ($\approx$ 9 nm RMS) of III-V semiconductors on Si. In comparison, the T1 detector of the microscope
*Contact author: yoan.leger@insa-rennes.fr

is less sensitive to the topology since its in-lens construction limits direct shadowing, while it is strongly sensitive to orientation contrast as demonstrated in part III.A. This makes the analysis of T1 images in the presence of surface roughness or features not straightforward as shown in Figure 6(c) and (d). On the contrary, a mixed image of the ETD and T1 images taken simultaneously, where the ETD signal is coded on the value (grey scale) and the T1 signal coded on the hue (color) allows a direct identification of the APDs as presented in Figure 6(e) and (f). In these examples, we confirm the direct correlation of the surface topology to the domain distribution, already widely discussed in the literature [51]. Thus, even in the unpolished, rough samples, APDs can be identified using DOCI.

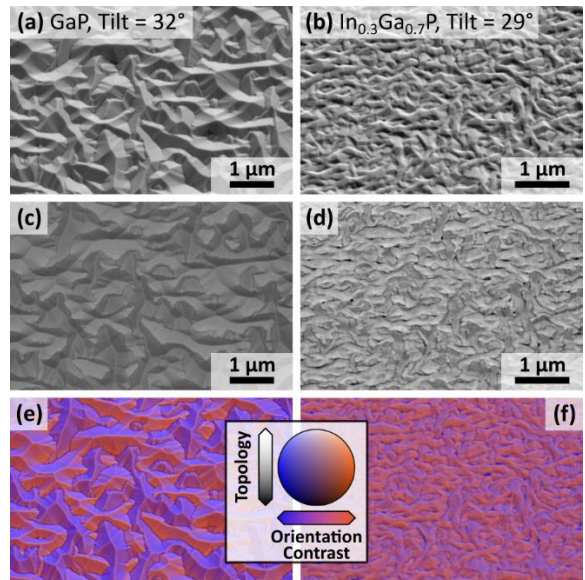

FIG 6. GaP/Si (a,c,e), $In_{0.3}Ga_{0.7}P$/SiGe/Si (b,d,f) using ETD (a and b), T1 (c and d) and mix of both (e and f), 2keV, 200pA, tilted at 32° for GaP and 29° for $In_{0.3}Ga_{0.7}P$.

### 3. Quantitative statistical analysis of APDs and APBs distributions in polished GaP/Si

To illustrate the potential of the DOCI technique for extracting statistical data from samples, a complete image analysis was performed on a GaP/Si sample polished by CMP. DOCI of GaP/Si was performed at 31°, and a tilt correction applied to correct the deformations due to tilting the sample (Figure 7. (a)). A multistep binarization (see Supplementary Materials) was then applied to extract the APD distribution (Figure 7(b)) and APB distribution (Figure 7(c)) at the sample surface. From these binarized images, a number of statistical information can be extracted. From the domain analysis (Figure 7(b)), the direct analysis of the image reveals an average surface polarity $P_{APD} = (S_w - S_b)/(S_w + S_b) = 0.015$, where $S_{w(b)}$ refers to the total surface of the white (black) phase domains. This means that the surface is almost evenly shared between the two polarities [9]. The radial Auto Correlation Function (ACF) $G_r$ is also extracted from Figure 7(b) and plotted in Figure 7(d). Note that in this domain statistical analysis, a 360°-average is considered, thus hiding any influence of anisotropic inhomogeneities of the domains. While such analysis is feasible with DOCI, it went beyond the scope of the present work. The analysis of $G_r$ local maxima allows the determination of the mean distance between two domains of the same phase, separated by one domain of opposite phase $L_{APD} = 0.8$ µm. An exponential fit of the $G_r$ ACF, also allows the determination of the correlation length. The characteristic length, i.e. the smallest statistically-relevant characteristic domain size measured on the image, is $L_c = 116$ nm here. Finally, the boundaries binarization provided in Figure 7(c) can also be analyzed. A Hough transform is applied, that converts detected APBs into segments, from which lengths and orientations are extracted. From the analysis, the density of emerging APBs can finally be represented as a function of their in-plane orientation in Figure 7(e). The directions <110> and <100> identified experimentally directly from the wafer are written on a blue background at the top. The spatial distribution of the emerging APB density shows that [110], [1-10], as well as [100] and [010] are directions belonging to many emerging APBs. But the distribution also reveals that emerging APBs also appear to be more numerous along other intermediate angles. For comparison, the angles corresponding to some low-index in-plane directions have been reported in Fig. 7(e) (with a green background). While the origin of APBs formation was found to be mainly related to growth kinetics and coalescence in previous studies [51,52], the present statistical analysis shows unambiguously that, during their propagation, the atomic structure of APBs is not random, and their atomic configuration (interfacial reconstruction) probably driven by different charges sharing with ladder and zigzag atomic configurations [25].

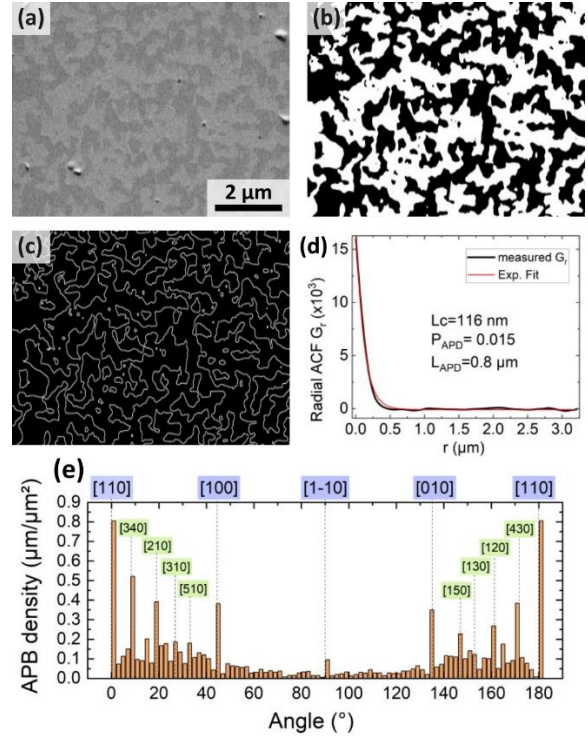

FIG 7. (a) Direct Orientation Contrast Imaging of the two opposite APDs with tilt correction. (b) Multistep thresholding binarization of (a). (c) Edge detection of (a) highlighting the existing APBs. (d) APD grain analysis by applying radial autocorrelation function on (b). (e) Determination of APBs densities with respect to angle after Hough transform done on (c).

## IV. CONCLUSIONS

DOCI is introduced as a promising SEM technique, using in-lens detector, that allows nondestructive, fast, more affordable direct imaging and characterization of APDs on zinc-blende III-V materials and appears as a relevant preliminary method for in-depth analyses such as EBSD and TEM with complementary capabilities. It is compatible with cross-sectional analysis and requires little to no surface preparation especially when the Z contrast is important between III and V elements. Under the numerous appropriate imaging conditions (tilt angle, beam energy, …), for III-V on non-polar substrate, it enables quantitative statistical analysis of APDs and APBs when samples are polished. On unpolished samples, it can be a rapid way to control APDs growth in the context of APDs burying. As for its use for OP-III-V, after regrowth and CMP, it allows surface measurement of domain size. Thus, when combined with standard SEM imaging before regrowth, it allows characterization of domain verticality. DOCI is expected to be a relevant method

*Contact author: yoan.leger@insa-rennes.fr

in imaging ADPs on other non-centrosymmetric materials.

## ACKNOWLEDGMENTS

The authors acknowledge RENATECH for technological support. This research was supported by "France 2030" with the French National Research agency OFCOC project (ANR-22-PEEL-0005), by the French National Research agency PIANIST project (ANR-21-CE09-0020), NUAGES project (ANR-21-CE24-0006) and NANOFUTUR project (ANR-21-ESRE-0012), and by the EUROPEAN Union (ERC-2022-COG, PANDORA, 101088331). This publication is supported by the European Union through European Regional Development Fund (ERDF), Ministry of Higher Education and Research, Brittany region, Conseil Départemental d'Ille-et-Vilaine, Rennes Métropole, through the CPER Project PhotBreizh.

*Contact author: yoan.leger@insa-rennes.fr

*Contact author: yoan.leger@insa-rennes.fr

*Contact author: yoan.leger@insa-rennes.fr